# Alignment-free dispersion measurement with interfering biphotons

Arash Riazi[1,*], Eric Y. Zhu[1], Changjia Chen[1], Alexey V. Gladyshev[2], Peter G. Kazansky[3], and Li Qian[1]

[1]Electrical and Computer Engineering Department, University of Toronto, 10 King's College Rd., Toronto, Canada M5S 3G4
[2]Fiber Optics Research Center, Russian Academy of Sciences, 38 Vavilov Street, 119333 Moscow, Russia
[3]Optoelectronics Research Centre, University of Southampton, Southampton SO17 1BJ, United Kingdom
*Corresponding author: arash.riazi@mail.utoronto.ca

**Measuring the dispersion of photonic devices with small dispersion-length products is challenging due to the phase-sensitive, and alignment-intensive nature of conventional methods. In this letter, we demonstrate a quantum technique to extract the second-, and third-order chromatic dispersion of a short single-mode fiber using a fiber-based quantum nonlinear interferometer. The interferometer consists of two cascaded fiber-based biphoton sources, with each source acting as a nonlinear beamsplitter. A fiber under test is placed in between these two sources, and introduces a frequency-dependent phase that is imprinted upon the biphoton spectrum (interferogram) at the output of the interferometer. This interferogram contains within it the dispersion properties of the test fiber. Our technique has three novel features: (1) The *broadband* nature of the biphoton sources used in our setup allows accurate dispersion measurements on test devices with *small dispersion-length* products; (2) our all-fiber common-path interferometer requires no beam alignment or phase stabilization; (3) multiple phase-matching processes supported in our biphoton sources enables dispersion measurements at different wavelengths, which yields the third-order dispersion, achieved for the first time using a quantum optical technique.**

Chromatic dispersion is an important physical property that affects the propagation of optical pulses in photonic systems; in linear systems it is used for pulse shaping [1], while in nonlinear systems it plays an important role in soliton propagation, and affects the efficiency of many nonlinear interactions [2]. Dispersion characterization is then crucial for designing optimized photonic devices. Significant effort has been expended in past decades on extracting the chromatic dispersion of materials using classical light sources. Techniques such as time-of-flight [3], and modulation phase shift [4] were introduced to measure the dispersion of components with large dispersion-length products. Components with small dispersion-length products were characterized with temporal [5], and spectral [6] white-light interferometry (WLI), with the latter proving to be the more robust method against environmental noise [7].

Spectral WLI can be subdivided into two categories: balanced and unbalanced. Balanced spectral WLI [6,8], which utilizes a balancing arm to cancel the effect of the first-order dispersion of the specimen, is capable of directly measuring the second-order dispersion [group velocity dispersion (GVD)] with a low-resolution spectrometer. The major drawback of this approach is that it requires a different "balancing arm length" for each wavelength at which the dispersion is measured. Moreover, the stability of the interferometer must be maintained with a precision on the order of a fraction of a wavelength. Unbalanced spectral WLI [9] addresses the alignment issue, and is considered to be a faster approach compared to balanced spectral WLI. However, it requires high spectral resolution [9,10] as well as a phase-stable interferometer. In addition, the GVD can only be obtained indirectly with unbalanced WLI; the approach may also suffer from errors arising from different curve-fittings approaches used for dispersion extraction [7,11].

Meanwhile, the potential advantages of quantum optics in metrology has led to the adaptation of some of the techniques mentioned above with nonclassical states of light such as entangled photons [12], and Fock states [11]. For example, an approach similar to the time-of-flight method [3] was used with energy-time entangled photons [12] to measure the second-order chromatic dispersion (GVD) of an optical fiber. However, just like its classical counterpart, the method required kilometers of fiber (large dispersion-length product) for an accurate estimation of the dispersion parameters.

In another scheme [11], a quantum *Mach-Zehnder* interferometer utilizing Fock states was used to measure the dispersion, albeit at only one wavelength, with a precision almost twice that of the classical balanced WLI. Aside from its promising precision, the second-order dispersion was directly obtained, even though the interferometer was unbalanced; this was due to the frequency correlation of the biphotons used in the interferometer. However, the technique still required the alignment of a *Mach-Zehnder* interferometer for both spatial, and polarization modes prior to the measurement. Additionally, active phase stabilization of the interferometer's reference arm was required.

In this letter, we overcome all the above issues by utilizing biphotons in a common-path, alignment-free nonlinear interferometer [13], and directly measure the second-, and the third-order dispersion (dispersion slope) of a short silica fiber. The interferometer we use is an all-fiber configuration consisting of two coherently-pumped fiber-based biphoton sources [14,15] [Fig. 1(a)]. The biphotons are generated through spontaneous parametric down-conversion (SPDC) in two periodically-poled silica fibers (PPSFs) [15], fibers with non-zero second-order

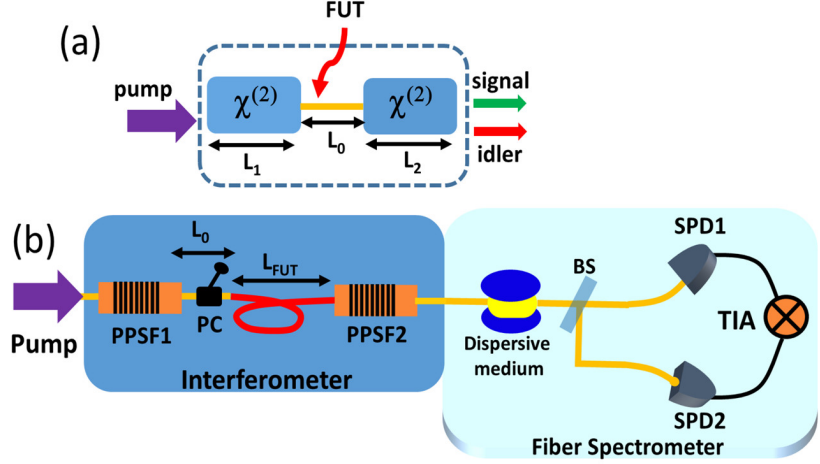

**Fig. 1.** a) The general scheme of our nonlinear interferometer, consisting of two periodically-poled silica fibers (PPSF) used as SPDC sources, and a drop-in fiber-under-test (FUT) of unknown dispersion. b) Experimental setup; two fiber-pigtailed PPSFs are used in our interferometer as the nonlinear media. Each PPSF generates biphotons in the 1.5-micron telecom region with a bandwidth of ~12 THz (100 nm). A polarization controller (PC) is used to send the proper pump polarization into the second PPSF. A dispersive medium (20 km of Corning SMF-28), a beam-splitter (BS), and a pair of single photon detectors (SPDs, IDQ ID220) are used as a fiber spectrometer. The signal and idler's arrival times are measured with a time interval analyzer (TIA, PicoQuant HydraHarp 400).

---

nonlinearities. The PPSFs are quasi-phase-matched [16] for the generation of biphotons at ~ 1550 nm, which interfere at the output of the interferometer. As the interferometer is common-path and all components support only one transverse spatial mode, our setup is alignment-free and inherently phase-stable. This feature is particularly important for quantum metrology due to the long integration times often required in the measurements. It is also worth mentioning that the frequency correlation of the biphotons removes the effects of all odd-order dispersion terms, which means the first-order dispersion (group delay) is automatically balanced out. Note that this feature would have not been possible in previous classical approaches [6,8].

We now elaborate further on the mechanism behind our nonlinear interferometer. The phase of the biphotons generated from the first SPDC source in Fig. 1(a) is modified by the fiber under test (FUT) placed in the middle section. These biphotons then interfere with the ones generated from the second SPDC source, and the resulting spectral interferogram reveals the dispersion information of the FUT. Two criteria must be met for biphoton interference: (1) The coherence length of the pump should be much longer than the optical path length difference between pump and signal/idler in the middle section [17,18], and (2) the frequency-dependent loss of the FUT at the signal (idler) frequency $\omega_s$ ($\omega_i$) should be negligible. When both conditions are met, the joint spectral intensity (JSI) of the biphotons, $S$, at the output of the interferometer can be expressed as [17,19]:

$$S(\omega_s,\omega_i) \propto |F(\omega_s,\omega_i)|^2 \{1 + \cos[\Phi(\omega_s,\omega_i)]\}, \qquad (1)$$

where $|F(\omega_s,\omega_i)|^2$ is the JSI of the biphotons generated from each biphoton source. Note that in deriving Eq. (1), we have assumed that the two biphoton sources are identical. As Eq.(1) indicates, the frequency-dependent phase $\Phi(\omega_s,\omega_i)$ introduced by the middle section of the interferometer results in the interference of the two biphoton states.

The pump used here will have a narrow linewidth due to its long coherence length, and its frequency, $\omega_p$, can be considered to be constant. Therefore, due to energy conservation we have $\omega_p = \omega_s + \omega_i$ for the frequency-conjugate pairs (signal and idler). For convenience in notation, we replace $\omega_s$ and $\omega_i$ with $\omega$ and $\omega_p - \omega$, respectively, and rewrite all the terms in Eq.(1) in terms of $\omega$.

The phase argument of the cosine function in Eq. (1) can be expressed as the sum of two components: $\Phi(\omega) = \Phi_{int}(\omega) + \Phi_{FUT}(\omega)$, where $\Phi_{int}(\omega)$ is the phase already present in the middle section before the FUT is inserted, while $\Phi_{FUT}(\omega)$ is the phase contribution of the FUT; $\Phi_{FUT}(\omega)$ can be further broken down into

$$\Phi_{FUT}(\omega) = \Delta k_{FUT} L_{FUT} = [k_{FUT}(\omega_p) - k_{FUT}(\omega) - k_{FUT}(\omega_p - \omega)] L_{FUT}. \qquad (2)$$

The Taylor expansion of Eq. (2) about $\omega_{deg} = \frac{\omega_p}{2}$ yields

$$\Phi_{FUT}(\omega) = \left[k_{FUT}^{(0)}(2\omega_{deg}) - 2k_{FUT}^{(0)}(\omega_{deg})\right]L_{FUT} - k_{FUT}^{(2)}(\omega_{deg})(\omega-\omega_{deg})^2 L_{FUT} + O\left[(\omega-\omega_{deg})^4\right], \quad (3)$$

where $k_{FUT}^{(N)}(\omega') = \frac{d^N k}{d\omega^N}\Big|_{\omega'}$. Note that higher-order dispersion terms of the pump [$k_{FUT}^{(N)}(\omega_p)$] are not present as the pump is continuous-wave (cw) and its linewidth (< 1 MHz) is much narrower than the biphoton spectrum. Equation (3) suggests that we can directly obtain the GVD, $k_{FUT}^{(2)}(\omega_{deg})$, by measuring the phase introduced by the FUT [$\Phi_{FUT}(\omega)$]. We also note that, in contrast to the classical balanced spectral WLI [6], the common procedure of balancing the test and the reference arms is automatically obviated due to the frequency-conjugate nature of the biphotons. Furthermore, since our interferometer is common-path, both pump and downconverted photons travel in a single spatial mode; this removes the requirement for phase stabilization, an otherwise essential requirement in both classical [6], and quantum WLI [11] schemes.

The experimental setup for dispersion measurement is shown in Fig. 1(b). A Toptica DL Pro is used as the pump in our experiment. The coherence length of the pump is significantly longer (~1 km) than the optical path length difference between pump and signal and idler (~ 3 cm), guaranteeing spectral interference at the output of the interferometer. The JSI [$S(\omega)$ in Eq. (1)] of the biphotons is measured with a fiber spectrometer [20][see Fig. 1(b)]; the biphotons at the output of the interferometer are sent to a dispersive medium, which maps their frequencies onto the temporal domain [21]. After time-tagged detection with single photon detectors, the spectrum is recovered by re-mapping the coincidence timing delays back into frequency. The minimum spectral resolution of our technique is limited primarily by the spectral resolution of the spectrometer, with the linewidth of the pump playing a secondary role. The minimum resolution of the spectrometer itself depends on the timing jitter of the single photon detectors (~ 256 ps), and the dispersion-length product of the dispersive medium (~ 340 ps/nm) [see Fig.2(b)] [20] used, which in our case gives us a resolution of 0.75 nm (~100 GHz). The pump linewidth begins to affect the spectral resolution as it becomes comparable to the spectral resolution of the spectrometer; however, the pump we used here has a long coherence length, which means its linewidth (~ 100 kHz) is negligible compared to the minimum resolution of the spectrometer (~ 100 GHz).

We use two fiber-pigtailed PPSFs as our broadband biphoton sources. They have identical fiber geometries, and composed of identical materials. Both PPSFs support two different phase-matched SPDC processes: (1) Type-0 SPDC when pumped at $\lambda_{deg}^{type-0}/2 = 780.2$ nm, and (2) type-I SPDC when pumped at $\lambda_{deg}^{type-I}/2 = 776.2$ nm. Figure 2 shows the JSIs of the biphotons generated from the two PPSFs when they are pumped at 780.2 nm (type-0 SPDC). We note that, in the 10 THz (~80 nm) region surrounding the degeneracy point, the two JSIs overlap well; this bandwidth is enough to extract the GVD of a photonic medium with dispersion-length product as small as ~0.002 ps/nm (see [22]), equivalent to 10 cm of SMF-28. The JSI for type-I SPDC is similar to what is observed in Fig. 2.

We first pump the interferometer at 780.2 nm for type-0 SPDC, and extract the JSI of the biphotons at the output of the interferometer without the FUT present. Since both PPSFs are fiber-pigtailed, $\Phi_{int}(\omega) \neq 0$ and the interference appears in the spectrum of the biphotons [Fig. 3(a)]. Note that the high fringe visibility (~75%) near the degeneracy point is due to the similarity of the two JSIs (Fig. 2).

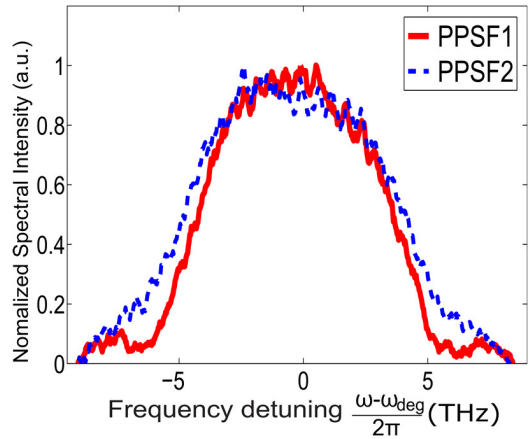

**Fig. 2.** The biphoton spectrum, $|F(\omega)|^2$, of each individual PPSF for type-0 SPDC phase-matching. There is a ~10 THz (80nm) window in which the emission spectra of the two PPSFs overlap well.

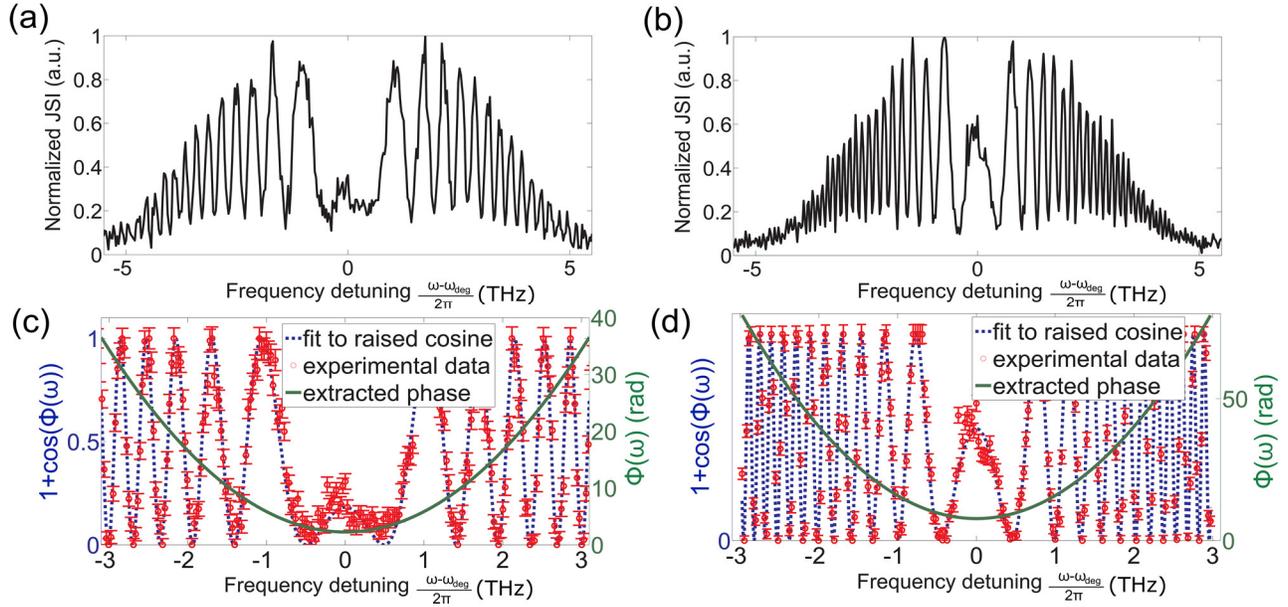

**Fig. 3.** The biphoton spectrum when the interferometer is pumped at 780.2 nm for type-0 SPDC a) without FUT, and b) with FUT in the middle section. A raised cosine function is fitted to each of the normalized biphoton spectra, and $\Phi(\omega)$ is extracted for type-0 SPDC c) without and d) with the FUT present in the interferometer.

After the FUT (5m of SMF-28) is placed inside the interferometer, the fringe spacing changes [Fig. 3(b)]. We then normalize the JSIs by the biphoton spectrum of each individual PPSF and fit the resultant functions to a raised cosine [see Fig. 3(c, d)]. This way we can extract the frequency-dependent phases $\Phi_{int}(\omega)$, and $\Phi_{int}(\omega) + \Phi_{FUT}(\omega)$. Now, by subtracting the two phase functions [green traces in Fig. 3(c, d)], we can obtain $\Phi_{FUT}(\omega)$ [Fig. 3(a)] and extract the GVD, $k^{(2)}_{FUT}(\omega_{deg})$. Then, we determine the dispersion parameter, $D(\lambda_{deg}) = -\frac{2\pi c}{(\lambda_{deg})^2} k^{(2)}_{FUT}(\omega_{deg})$ (ref. [1]), for the FUT at $\lambda_{deg}^{type-0} = 1560.4$ nm as 16.69(5) ps/nm/km. Now, by tuning the pump wavelength to 776.2 nm for type-I SPDC process, we can extract the dispersion parameter of the FUT at $\lambda_{deg}^{type-I} = 1552.4$ nm as well. Using the same procedures described above, we obtain $D(\lambda_{deg}^{type-I}) = 16.06(11)$ ps/nm/km [Fig. 3(b,c)]. Given the dispersion parameter at two different wavelengths, we can obtain the third-order dispersion for this wavelength range as 0.078(13) ps/nm²/km. We remark that the second- and third-order dispersion parameters measured here are both consistent with the specifications of SMF-28 (ref. [23]) [Fig. 4(c)].

It is worth mentioning that when the pump energy is drastically increased, our single-photon-level interferometer transforms into the high-gain nonlinear interferometer. In that case, our setup [Fig. 1(b)] resembles the so-called SU(1,1) interferometer [24] often used for low-noise phase measurements [13]. However, due to the high-gain nature of the parametric process [25], the number of photons per frequency mode in this regime is much larger; however, the two parametric down-converted beams from each nonlinear medium still interfere [26,27], resulting in spectral fringes (analogous to the spontaneous regime) that can be detected classically at the output of the interferometer.

In both the SPDC and high-gain regimes, the minimum dispersion-length product that can be measured is inversely proportional to the bandwidth of the spectral interference [see [22], and Eq. (3)]. With appropriate phase-matching [15], the spontaneous emission bandwidth of the nonlinear medium can be significantly increased, and one can measure the dispersion of devices with dispersion-length products well below 0.002 ps/nm. Meanwhile, the emission bandwidth in the high-gain regime increases proportionally with the parametric gain [28,29]; however, the effective bandwidth over which the spectral interference takes place remains unchanged [26] due to spectral-spatial correlations [30]. This effect can limit the interference bandwidth in the stimulated regime, and consequently constrains the minimum measurable dispersion-length product.

Future work could investigate the regime in which our nonlinear interferometer is seeded with a classical signal beam from, for example, a tunable laser source. This would be similar to the stimulated emission tomography technique [31], and would allow for a fully classical dispersion measurement with our current setup.

In summary, we have introduced a common-path nonlinear interferometer, and used it for a broadband phase measurement. Specifically, we demonstrated that the phase imprinted upon the spectral interferogram can be used to extract the second-, and third-order dispersion of a single mode fiber embedded within the interferometer. The collinear broadband emission of the biphoton sources and the common-path nature of our interferometer makes it inherently phase-stable, robust, and capable of measuring the dispersion of materials with small dispersion-length products. The biphoton sources we used here are best suited for the dispersion measurement of specialty fibers, where off-the-shelf components can be connected and coupled with minimal effort. However, our method is source-independent, and can be implemented with other waveguide-based biphoton sources.

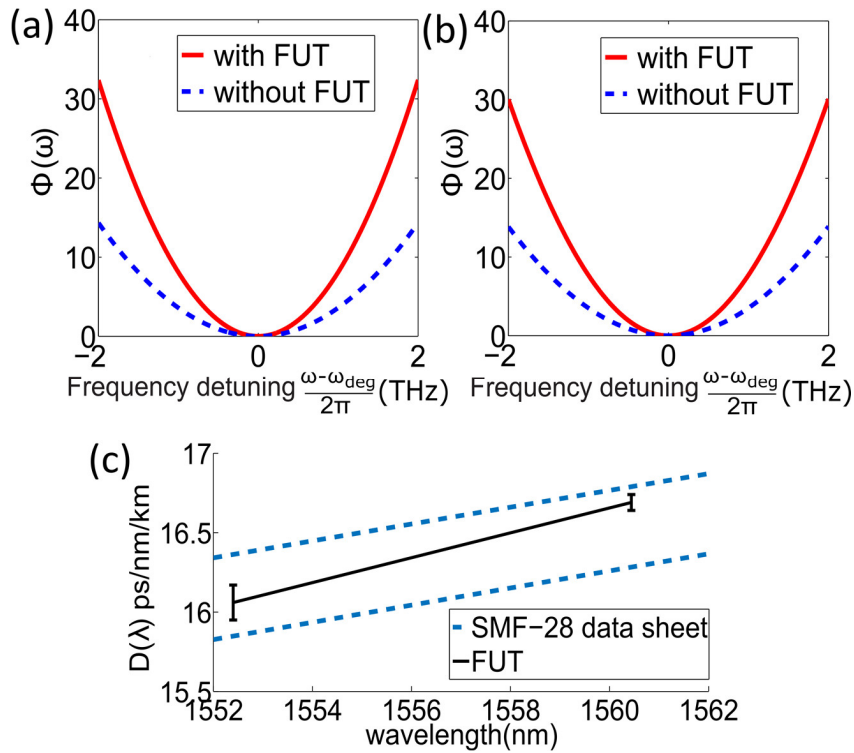

**Fig. 4.** Extracted interferometric $\Phi(\omega)$ with (solid line, red) and without (dashed line, blue) the FUT present for a) type-0 and b) type-I SPDC. The offset phase at degeneracy has been zeroed. C) The second-order dispersion parameter (solid line) of a 5 m Corning SMF-28 patchcord extracted at $\lambda_{deg}^{type-0}$ and $\lambda_{deg}^{type-I}$. The dashed line represents the bounds for the dispersion parameter of a typical SMF-28 patchcord, with a zero-dispersion slope of 0.085ps/(nm²km) and zero-dispersion wavelengths of 1308 nm (upper bound) and 1318 nm (lower bound).

**Funding.** Natural Sciences and Engineering Research Council of Canada (RGPIN-2014-06425, RGPAS 462021-2014).
**Acknowledgment.** The authors would like to thank Prof. J. E. Sipe for fruitful discussions. We would also like to thank the anonymous reviewer for suggesting the seeded nonlinear interferometer idea.